\documentclass[aps,prl,twocolumn,showpacs,superscriptaddress,amsmath,longbibliography,amssymb]{revtex4-1}
\usepackage[utf8x]{inputenc}
\usepackage{graphicx}
\usepackage{dcolumn}
\usepackage{bm}
\usepackage{float}
\usepackage{amsmath}
\usepackage{color}

\begin{document}

\title{The cell adaptation time sets a minimum length scale for patterned substrates}

\author{Diogo E. P. Pinto}
\affiliation{Departamento de F\'{i}sica, Faculdade
de Ci\^{e}ncias, Universidade de Lisboa, Campo Grande, P-1749-016 Lisboa,
Portugal}
\affiliation{Centro de F\'{i}sica Te\'{o}rica e Computacional, Campo
Grande, P-1749-016 Lisboa, Portugal}

\author{Gonca Erdemci-Tandogan}
\affiliation{Department of Physics, Syracuse
University, Syracuse, New York 13244, USA}
\affiliation{Institute of Biomaterials and Biomedical Engineering, University
of Toronto, Toronto, ON, M5S 3G9, Canada}

\author{M. Lisa Manning}
\affiliation{Department of Physics, Syracuse
University, Syracuse, New York 13244, USA}

\author{Nuno A. M. Araújo}
\affiliation{Departamento de F\'{i}sica, Faculdade
de Ci\^{e}ncias, Universidade de Lisboa, Campo Grande, P-1749-016 Lisboa,
Portugal}
\affiliation{Centro de F\'{i}sica Te\'{o}rica e Computacional, Campo
Grande, P-1749-016 Lisboa, Portugal}

\begin{abstract}
The structure and dynamics of tissue cultures depend strongly on the physical
and chemical properties of the underlying substrate. Inspired by previous
advances in the context of inorganic materials, the use of patterned culture
surfaces has been proposed as an effective way to induce space-dependent
properties in cell tissues. However, cells move and diffuse and the transduction of
external stimuli to biological signals is not instantaneous. Here, we show
that the fidelity of patterns depends on the relation between the diffusion
($\tau_D$) and adaptation ($\tau$) times. Numerical results for the
self-propelled Voronoi model reveal that the fidelity decreases with
$\tau/\tau_D$, a result that is reproduced by a continuum reaction-diffusion
model. We derive a minimum length scale for the patterns that depends on
$\tau/\tau_D$ and can be much larger than the cell size.
\end{abstract}

\maketitle

\section{Introduction}
The regulated growth and maintenance of a living tissue under controlled
conditions is a major challenge for cell biology and tissue engineering. The
standard procedure consists in the use of culture surfaces to support and
guide the cells~\cite{Lo2000,Griffith2002,Rabodzey2008,Tambe2011,Park2015}.
An extensive body of research shows that the cell morphology and dynamics are
sensitive to the physical and chemical properties of the
substrate~\cite{Lo2000,Discher2005,Guo2006,Neuss2009,Tambe2011,Murrell2011,Song2013,Sunyer2016,Janmey2019}.
For example, it has been shown that substrate stiffness can significantly
affect the geometry of cultured cells, including their spreading
area~\cite{Yeung2005,Janmey2019}, volume~\cite{Guo2017}, and shape
elongation~\cite{Devany2019}. In addition, the nanotopography of the substrate
can alter cell polarization, shape, and
motility~\cite{Davis2011,Tseng2013,Jeon2015,Mengsteab2016}. Thus, the effort
has been in the design of biocompatible substrates that regulate the
individual and collective dynamics of cells. 

There is a sustained interest in the possibility of generating spatial
patterns of cells with different properties, which is critical for
morphogenesis, collective cell motion, and wound
healing~\cite{Nelson2005,Richert2008,Davis2011,Tseng2013,Jeon2015,Mengsteab2016,
Horejs2018,PerezGonzalez2018,Brasch2019}. In development, the processes that
typically generate two tissue types separated by a boundary have been studied
extensively~\cite{Steinberg1963,Harris1976,Brodland2002,Foty2005,Amack2012},
and often occur because two different cell types, through various mechanisms,
prefer to be surrounded by cells of the same
type~\cite{Maitre2012,Krieg2008,Manning2010,Sussman2018a,Sahu2020}.  However,
in vitro, an alternative approach is to culture a single cell type on a
patterned substrate, and allow the patterned substrate to change the
properties of cells to generate a
pattern~\cite{Jeon2015,Richert2008,Mengsteab2016,Davis2011}.
 
Patterned substrates have been used to a large extent in the context of
inorganic materials~\cite{Kumacheva2002,Fustin2003,Dziomkina2005,Cadilhe2007}.
However, their use for biological systems raises several additional
difficulties. Besides the need for biocompatible materials, the transduction
of external stimuli into biological signals that control the cell morphology
and mechanics is not instantaneous. It requires a hierarchy of biochemical
processes, which sets a characteristic adaptation time that can extend over
hours~\cite{Ebara2015}. The problem is that, within the adaptation time scale,
cells might move around and explore other regions of the substrate. Thus, the
fidelity of patterns in the regulation of cell tissues should depend on how
the adaptation time compares with the other relevant time scales. This is
precisely what we study here.

We consider an epithelial confluent tissue on a simple patterned substrate,
consisting of two halves that solely differ in the cell-substrate interaction
(see Fig.~\ref{cellscheme}). We describe the tissue with the self-propelled
Voronoi model, where the cell-substrate interaction is included in the
preferential geometry of each cell, as cell shapes change as a function of
substrate properties~\cite{Yeung2005}, and cell shape in turn governs the rate
of cell diffusion in monolayers~\cite{Park2015}. We show that the fidelity of
the pattern in the regulation of the tissue properties is compromised
significantly when the adaptation time competes with the diffusion time of
cells.

\section{Model}
\begin{figure}
\includegraphics[width=\columnwidth]{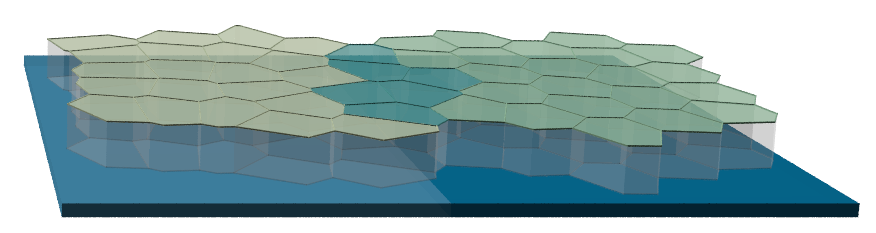}
\caption{\label{cellscheme} Schematic representation of the system. We
consider a confluent tissue on a squared substrate with two regions of equal
linear length $L/2$, which differ in the target value of the shape index $p_0$
of cells: $p_0=p_A$ on the brighter side (left) of the substrate and $p_0=p_B$
on the darker one (right), with $p_B>p_A$. The color of the cells is related
to their actual shape index $p_0(t)$, which is equal to $p_A$ for the ones in
the left and $p_B$ for the ones in the right. The darker cells (in the middle)
have an intermediate value of $p_0(t)$, i.e., $p_A<p_0(t)<p_B$.}
\end{figure}
We model the confluent tissue as a monolayer of $N$ cells using the
self-propelled Voronoi model~\cite{Bi2016,Bi2014,Bi2015}. Each cell $i$ is
represented by its center $\bm{r}_i$ and its shape is given by the Voronoi
tesselation of the space. The stochastic trajectories of cells are obtained
from a set of Langevin equations of motion,
\begin{equation} \label{motion}
\frac{d\bm{r}_{i}}{dt}=\mu \bm{F}_{i}+v_{0}\hat{\bm{n}}_{i} ,
\end{equation}
where $\bm{F}_{i}$ is the net force acting on cell $i$, $\mu$ is the mobility
of the cell, $v_{0}$ the self-propulsion speed, and
$\hat{\bm{n}}_{i}=(cos\theta_{i},sin\theta_{i})$ is a polarity vector which
sets the direction of the self-propulsion force. For simplicity, we consider
that $\theta_{i}$ is a Brownian process given by,
\begin{equation}
\dot{\theta_{i}}=\eta_{i}(t), \hspace{0.1cm} \langle
\eta_{i}(t)\eta_{j}(t')\rangle = 2D_{r}\delta(t-t')\delta_{ij} , 
\end{equation}
where $\eta_{i}(t)$ is a uncorrelated random process of zero mean and its
variance sets the rotational diffusion $D_{r}$.

The net force $\bm{F}_{i}$ describes the multibody cell-cell interaction and
it is given by $\bm{F}_{i}=-\nabla E_i$, where $E_i$ is the energy
functional for cell $i$~\cite{Farhadifar2007,Fletcher2014},
\begin{equation}\label{eq::en.functional}
E_{i}=K_{A}[A_{i}-A_{0}]^2+K_{P}[P_{i}-P_{0}]^2 \ \ ,
\end{equation}
where $A_{i}$ and $P_{i}$ are the area and perimeter of cell $i$,
respectively, and $A_{0}$ and $P_{0}$ are their target values. The first term
accounts for the cell incompressibility and the resistance to height
fluctuations. The second term accounts for the active contractility of the
actomyosin subcellular cortex and effective cell membrane tension, due to
cell-cell adhesion and cortical tension. $K_A$ and $K_P$ are the area and
perimeter moduli. By rescaling energy in units of $K_AA_0^2$, we obtain four
adimensional quantities: two that characterize the area and the perimeter of
the cell ($a_{i}=A_{i}/A_{0}$ and $p_{i}=P_{i}/\sqrt{A_{0}}$), a shape parameter
$p_0=P_0/\sqrt{A_0}$, and $r=K_A A_0/K_p$ (see
\textit{Supplementary Information}). Without loss of generality, below, all
lengths are in units of $\sqrt{A_{0}}$ and time is in units of $1/(\mu
K_{A}A_{0})$. 

Cells diffuse with a diffusion coefficient $D$ that depends on the four
control parameters: the speed of self-propulsion $v_0$, the rotational
diffusion $D_r$, the shape index $p_0$, and the ratio $r$. For fixed values of
$v_0$ and $D_r$, as considered here, the model yields a rigidity transition at
a threshold value of the shape index $p_0=p_c$: from a solid-like state with
finite shear modulus, for $p_0<p_c$, to a fluid-like state of zero rigidity,
for $p_0>p_c$, where cell rearrangements are more
frequent~\cite{Bi2016,Merkel2019}.

We consider a squared substrate of linear length $L=\sqrt{N}$, where the value
of the target shape index ($p_0$) is spatially dependent. As schematized in
Fig.~\ref{cellscheme}, we split the substrate in half, with different values
of $p_0$ in each side. Thus, cells on the left-hand side have a target
$p_0=p_A$, while the ones on the right-hand side have $p_0=p_B$, where
$p_B>p_A$. When a cell diffuses from one side to the other, their target value
of the shape index in Eq.~\eqref{eq::en.functional} changes accordingly,
within a characteristic adaptation time $\tau$. Thus, we consider that the
time dependence of the shape index $p_{0,i}$ for cell $i$ is given by,
\begin{equation} \label{eq::tau}
\dot{p}_{0,i}(\Delta t_{i})=\frac{1}{\tau}[p_{0,i}(\infty)-p_{0,i}(\Delta
t_{i})] ,
\end{equation}
where $\Delta t_{i}$ is the time interval since the cell crossed the line
dividing the substrate, for the last time. $p_{0,i}(0)$ is the shape index of
cell $i$ before crossing and $p_{0,i}(\infty)$ is the target value in the new
side.

\section{Results}
To study the role of the adaptation time $\tau$, we first consider a pair of
values for the shape index for which the confluent tissue is in a fluid-like
state at both sides of the substrate: $p_A=3.875$ and $p_B=3.9$. For these
values, the cell diffusion coefficients on each side differ by less than
$15\%$: $D_A=3.61\times10^{-3}$ and $D_B=4.13\times10^{-3}$, obtained from the
mean squared displacement. Below, we assume $D_A=D_B=D$. 

Initially, all cells are fully adapted to the underlying substrate (see
\textit{Methods}). As time evolves, cells diffuse and cross from
one side to the other. However, due to the finite adaptation time, their
target shape index $p_{0,i}$ changes in time, as given by Eq.~\eqref{eq::tau},
and thus cells of different shape indices mix in both sides of the substrate.
To characterize this mixing, we measure the demixing
parameter, $D_{p}$, defined as, 
\begin{equation} \label{DPeq}
D_{p}=\frac{1}{N}\sum_{i}^{N}\frac{1}{N_{i, neigh}}\sum_{j}^{N_{i, neigh}}
H(\epsilon-|p_{0,i}-p_{0,j}|),
\end{equation}
where the out sum is over all cells and the inner sum is over the $N_{i,
neigh}$ neighbors of cell $i$~\cite{Sahu2020}. $H(\epsilon-|p_{0,i}-p_{0,j}|)$ is the
Heaviside step function and $\epsilon$ is a threshold that we set to $\epsilon
=10^{-5}$ (see \textit{Supplementary Information} for the dependence on
$\epsilon$). For $D_{p}=1$ the cells in the confluent tissue are completely
segregated by their shape index, whereas for $D_p=0$ they are fully mixed,
i.e., each cell is surrounded by cells with a different shape index.

\begin{figure*}[t]
\includegraphics[width=0.8\textwidth]{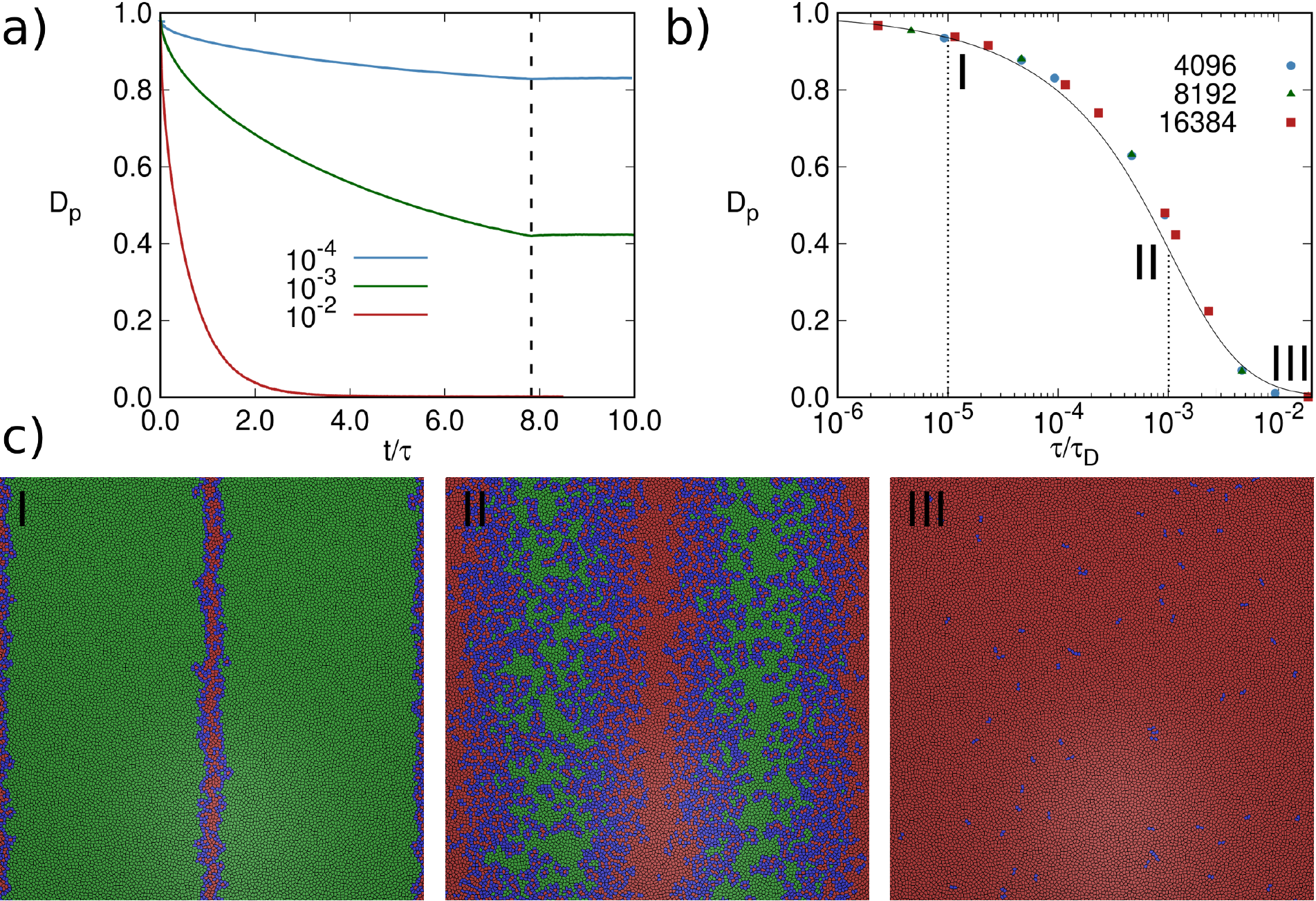}
\caption{\label{DP}Dependence of the demixing parameter $D_p$ on the two
relevant time scales: adaptation $\tau$ and diffusion $\tau_D$ times. (a) Time
dependence of the demixing parameter, where time is rescaled by the adaptation
time $\tau$. Different curves are for different values of $\tau/\tau_D$,
namely, $10^{-4}$, $10^{-3}$, and $10^{-2}$. The vertical dashed line
corresponds to $\ln[(p_B-p_A)/\epsilon]$, which is the time it takes for the
target shape index of a cell $i$ that crosses to the right-hand side, with
$p_{0,i}=p_B$ to become $p_{0,i}=p_A+\epsilon$, as given by
Eq.~\eqref{eq::tau}.(b) Demixing parameter as a function of $\tau/\tau_D$ for
different system sizes, where the number density of cells is kept constant to
unity, i.e., $L=\sqrt{N}$. The (black)-solid line is given by
Eq.~\eqref{DPanal}, derived from a continuum model, with
$\alpha=0.0866\pm0.0009$. (c) Snapshots of the confluent tissue obtained
numerically at time $10\tau$, for three different values of $\tau/\tau_D$,
namely, $10^{-5}$, $10^{-3}$, and $10^{-2}$. The color of each cell depends on
the demixing parameter: green ($D_p=1$), red ($D_p=0$), and blue ($0<D_p<1$).
It is clear that the cluster of red cells is formed around the line dividing
the substrate into two parts (see Fig.~\ref{cellscheme}) and it grows with
$\tau/\tau_D$ until it spans the entire tissue. Results in (a) and (b) are
averages over ten independent samples.} 
\end{figure*}
Figure~\ref{DP}(a) shows the time dependence of the demixing parameter for
different values of the adaptation time $\tau$, where time is rescaled by
$\tau$. As cells mix, $D_p$ decreases and saturates asymptotically. Different
curves are for different values of $\tau/\tau_D$, where $\tau_D=L^2/D$ is the
diffusion time. In Fig.~\ref{DP}(b) is the asymptotic value of $D_p$ as a
function of $\tau/\tau_D$ for different numbers of cells (same density). A
data collapse is observed, which shows that finite-size effects are
negligible. The monotonic decrease of the asymptotic value of $D_p$ with
$\tau/\tau_D$ hints at a competition between two time scales: the adaptation
and the diffusion time.  When the adaptation time is negligible
($\tau\lll\tau_D$), cells adapt rapidly to the underlying substrate, with
$D_p\approx1$. When the two time scales compete, the value of $D_p$ should
depend on the ratio between the two. In the limit where they are of the same
order, $D_p$ should vanish, for the cell changes sides before fully adapting
to the new shape index. Thus, large values of the adaptation time compromise
the control over the shape of the tissue boundaries via patterned substrates.

The demixing parameter is not uniformly distributed in space. In
Fig.~\ref{DP}(c) are three snapshots of the tissue obtained at time $10\tau$,
for different values of $\tau/\tau_D$. The color of cells depends on the value
of the demixing parameter. Cells that are surrounded by cells of the same
target shape index $p_0$ are in green ($D_p=1$), the ones surrounded by cells
of a different $p_0$ are in red ($D_p=0$). The ones with intermediate values
of $D_p$ are in blue. One clearly sees that the green cells are in the center
of each half, whereas red cells are concentrated around the boundaries: middle
and borders, due to the periodic boundary conditions.  However, the width of
the regions of green and red cells depends on the value of $\tau/\tau_D$. 

\begin{figure*}[t]
\includegraphics[width=0.8\textwidth]{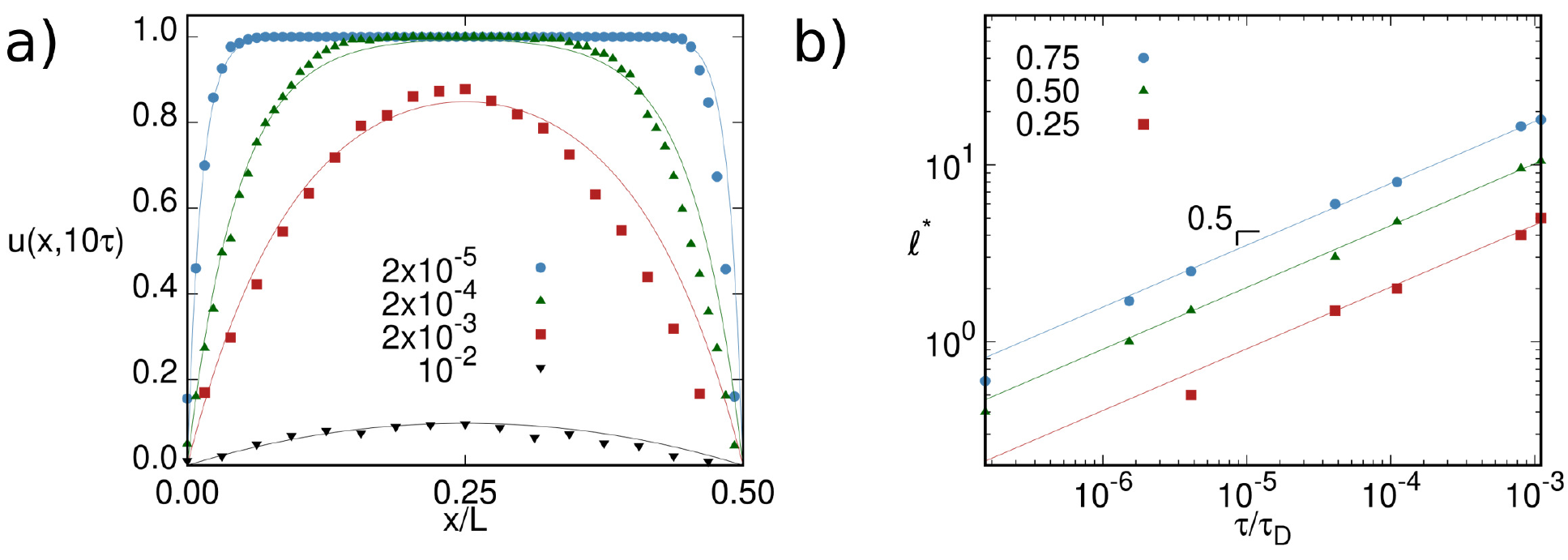}
\caption{\label{l} Spatial distribution of cells with $D_p=1$. (a) Profile
$u(x,t)$ of the fraction of cells with $D_p=1$, for different values of
$\tau/\tau_D$, where $x$ is the spatial coordinate along the horizontal
direction, $t=10\tau$, $\tau$ is the adaptation time, and $\tau_D$ the
diffusion time. The lines are given by Eq.~\eqref{eqss}, which is derived from
a continuum model, using $\phi$ as a fitting parameter. (b) Value of
$x=\ell^*$ at which $u(\ell^*,10\tau)$ is $0.25$ (squares), $0.50$ (triangles)
or $0.75$ (circles), as a function of $\tau/\tau_D$. Results are averages over
ten independent samples.}
\end{figure*}
We define $u(x,t)$ as the fraction of cells that are green ($D_p=1$) at time
$t$, where $x\in[0,L]$ is the spatial coordinate along the horizontal
direction. To compute $u(x,t)$ numerically, we divide the system into vertical
slices and measure the fraction of cells with $D_p=1$ within each slice.  The
results for $u(x,10\tau)$, for different values of $\tau/\tau_D$ are shown in
Fig.~\ref{l}(a), for $x\in[0,L/2]$. As suggested by Fig.~\ref{DP}(c), there
are more green cells at $x=L/4$ but, the fraction of cells and the width of
the profile decreases with $\tau/\tau_D$. The latter scales with
$\sqrt{\tau/\tau_D}$ as expected for a diffusive process (see
Fig.~\ref{l}(b)).

To describe the competition between cell diffusion and adaptation time, we now
propose a continuum model to describe the time evolution of $u(x,t)$. For
simplicity, we take advantage of the symmetry of the problem and only focus on
$x\in[0,L/2]$. We consider a reaction-diffusion equation for $u(x,t)$,
\begin{equation} \label{Eq}
u_{t}(x,t)=D^*u_{xx}(x,t)+T[1-u(x,t)] .
\end{equation}
where $u_t$ is the time derivative and $u_{xx}$ is the second space
derivative. The first term on the right-hand side is a diffusive term that
describes the collective diffusion of cells, with an effective diffusion
coefficient $D^*$. The second term is a reaction term, which describes the
adaptation of cells to the local environment. The adaptation is proportional
to the fraction of cells that are not adapted, i.e. $1-u(x,t)$, and occurs at
a rate $T$ that is proportional to the inverse of the adaptation time $\tau$.
Since we start from a demixed state, the initial conditions are $u(x,0)=1$ and
the boundary conditions are $u(0,t)=u(L/2,t)=0$ at all times. 

As derived in the \textit{Supplementary Information}, the control parameter for
the dynamics of the continuum model is the ratio $\phi^2=T(L/2)^2/D^*$. Since
$T\sim\tau^{-1}$ and $D^*\sim D$, then $\phi^2\sim L^2/\tau D=\tau_D/\tau$,
which is the ratio between the diffusion and adaptation time scales. We define
$\phi^2=\alpha L^2/\tau D$, where $\alpha$ is a prefactor that depends on the
geometry of the substrate and the value of the shape index on both sides.
Asymptotically, $u(x,t)$ converges to a stationary state $u_E(\hat{x})$,
\begin{equation} \label{eqss}
u_{E}(\hat{x})=\frac{1+e^{\phi}-e^{-\phi(\hat{x}-1)}-e^{\phi
\hat{x}}}{1+e^{\phi}} ,
\end{equation}
where $\hat{x}=2x/L$. As shown in Fig.~\ref{l}(a), this analytical solution
(solid lines) is in qualitative and quantitative agreement with the numerical
results for the self-propelled Voronoi model, where we set
$\alpha=0.0866\pm0.0009$ for all curves, obtained by a fit using the least squares method.

To compute the demixing parameter from the profile in the stationary state,
$u_E(x)$, we consider a mean-field approach, where the probability that two
neighboring cells are in the same state is $u_E^2(x)$ and so $D_p=\int_0^1
u_E^{2}(\hat{x})d\hat{x}$, which gives,
\begin{equation} \label{DPanal}
D_{p}= 1+\frac{1}{1+cosh(\phi)}-\frac{3 tanh(\phi/2)}{\phi} \ \ .
\end{equation}
This solution is the solid line in Fig.~\ref{DP}(b), which is in quantitative
agreement with the numerical results.

The numerical and analytical results suggest that the fidelity of a
patterned substrate in the control of the morphology of a tissue is
significantly dependent on the ratio between the diffusion and the adaptation
time. Ideally, full control would imply $D_p=1$. The lower is the value of
$D_p$, the less efficient is the use of a pattern. Let us define $\delta$ such
that a tissue with $D_p<\delta$ is considered mixed. Since $D_p$ increases
monotonically with $\phi$ (see Fig.~\ref{DP}(b)), we take the limit of
vanishing $\phi$ and $D_p$. From a Taylor expansion about $\phi=0$, we obtain
$D_p=\phi^4/120+O(\phi^5)$. Thus, there is a minimum length,
\begin{equation}\label{min.length}
L_\textrm{min}=\left[\frac{\tau
D}{\alpha}\sqrt{120\delta}\right]^{\frac{1}{2}} \ \ ,
\end{equation}
below which the cells in the tissue are mixed, which sets a lower bound for
the size of the patterns.

\begin{figure*}[t]
\includegraphics[width=0.8\textwidth]{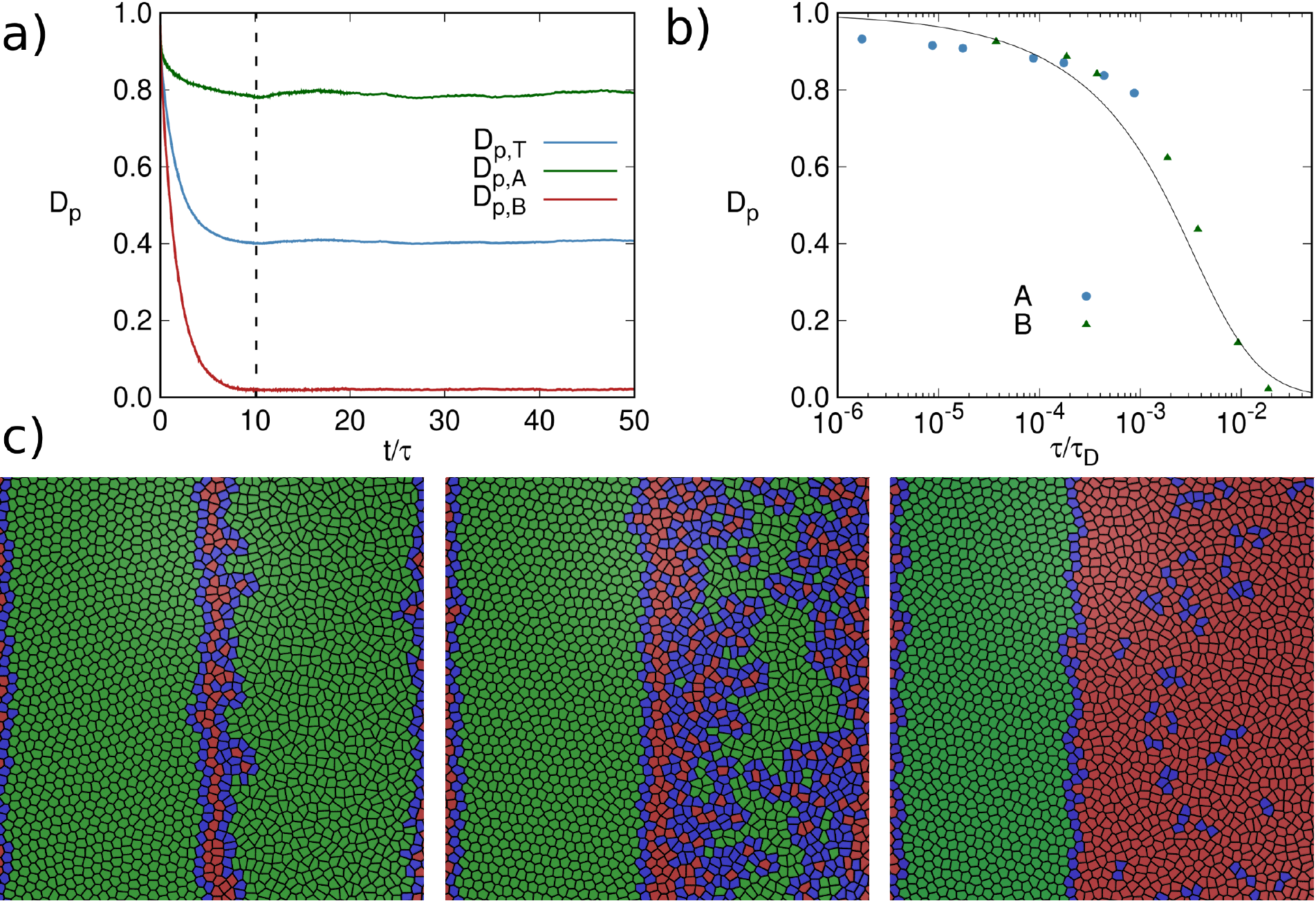}
\caption{\label{DP2}Dependence of the demixing parameter for the solid-fluid
case. (a) Time dependence of the demixing parameter, where time is rescaled by
the adaptation time ($\tau=5000$), for $N=2048$. Different curves are for the
side $A$, $B$, and both sides. (b) Demixing parameter as a function of
$\tau/\tau_D$ for both sides, obtained at $50\tau$. The (black)-solid line is
given by Eq.~\eqref{DPanal}, derived from the continuum model, with
$\alpha=0.276\pm0.003$. (c) Snapshots of the confluent tissue, obtained
numerically for different values of $\tau/\tau_D$, namely, $10^{-3}$,
$10^{-2}$, and $10^{-1}$. The color of each cell depends on the demixing
parameter: green ($D_p=1$), red ($D_p=0$), and blue ($0<D_p<1$).  Results in
(a) and (b) are averages over ten independent samples and the value of
$\tau_D$ is obtained for the liquid-like side.} 
\end{figure*}
So far, we considered a pair of target shape indices such that both sides are
in a fluid-like state. We study now the solid-fluid case, by setting
$p_A=3.65<p_c$ and $p_B=3.9$ (as before). Figure~\ref{DP2}(a) shows the time
dependence of the demixing parameter for the side $A$, $B$, and both sides.
Different from the fluid-fluid case, where the time dependence of $D_p$ was
similar for both sides, here we observe that $D_p$ vanishes for the
liquid-like side, whereas in the solid-like side it saturates at $\approx0.8$.
This break of symmetry is observed for a wide range of parameters, as seen in
Fig.~\ref{DP2}(b) from the dependence of the value of $D_p$ on the left- and
right-hand sides on $\tau/\tau_D$, where $\tau_D$ is the one for the
liquid-like state. For all values of $\tau/\tau_D$ the demixing parameter is
higher in the solid-like state than in the liquid one. This asymmetry stems
from the difference in the effective value of the diffusion coefficient $D$ in
both sides. For the solid-like state, $D\approx0$ and thus adaptation
is much faster than diffusion. Cells have enough time to adapt to the
new target shape index, which yields a high value of $D_p$ that does not
depend strongly on $\tau$ (see also snapshot for different values of $\tau$ in
Fig.~\ref{DP}(c)). By contrast, for the liquid-like state, the value of $D_p$
strongly depends on the value of $\tau$ as in the liquid-liquid case, see
Fig.~\ref{DP}. In fact, the dependence of $D_p$ on $\tau/\tau_D$ in
Fig.~\ref{DP2} for the liquid-like side is well described by
Eq.~\eqref{DPanal}, solid curve in Fig.~\ref{DP2}(b) (further results for
different substrates are discussed in the \textit{Supplementary Information}).

\section{Conclusion}
We included the adaptation time of cells to external stimuli in a minimal
model for confluent tissues. We found that the use of patterned substrates to
regulate the tissue properties is compromised significantly when the
adaptation time competes with the cell diffusion time. The latter depends on the
characteristic length of the pattern $L^*$. From a continuum description based
on a reaction-diffusion equation, we derived an analytic expression for the
minimum length $L_\textrm{min}$ for the pattern to be effective. For
$L^*>L_\textrm{min}$, cells have enough time to adapt to the local
cell-substrate interaction and the heterogeneous distribution of cell shapes
reproduces the symmetries of the pattern, with a clear segregation by shape
index. By contrast, for $L^*<L_\textrm{min}$, cells do not fully adapt to the
local cell-substrate interaction and their shape index depends on their
individual trajectories. 

For inorganic materials, the goal has been to reduce the length scale of the
patterns and achieve a control at the level of an individual
particle~\cite{Cadilhe2007,Araujo2008}. By contrast, in the case of cell
tissues, we show that the relevant length scale is set by the dynamics.
Experimentally, it was shown that cells cultured on a shape memory polymer
substrate take about $36$ hours to adapt to changes in the structure of the
substrate~\cite{Ebara2014}. If we consider a typical diffusion coefficient of
a cell in a confluent tissue of
$D\approx0.1\mu\text{m}^2\text{min}^{-1}$~\cite{Park2015}, from
Eq.~\eqref{min.length}, we obtain that $L_\textrm{min}\approx 55\mu\text{m}$,
which is roughly four times the size of a single cell. 

We considered a simple pattern but, it is straightforward to extend the
conclusions to other patterns. In fact, the competition between diffusion and
adaptation is so general that it should apply even to heterogeneous random
substrates. These substrates are usually characterized by a correlation length
$\xi$ that plays the role of $L^*$. So, only for $\xi>L_\mathrm{min}$, cells
are expected to segregate based on their shape index, as defined by the local
properties of the substrate.

The identification of the mechanisms responsible for the emergence of spatial
cell patterns in a developing organism has been a subject of intensive
research and discussion over the
years~\cite{Steinberg1963,Manning2010,Mertz2012,Maitre2012}. A recent study
combines theory and experiments to show that cell sorting and
compartmentalization in living organisms might be driven by surface tension
due to differential adhesion~\cite{Sahu2020}. However, the use of cell
mixtures in vitro encompasses multiple challenges, which include the lack of
control over the spatial distribution of cell types. We have shown that, above
a certain length scale, the spatial distribution of cell properties can be
controlled by the substrate pattern.

For simplicity, we assumed that the cell-cell and cell-substrate interactions
depend on the substrate but not on the cell itself. A recent study
shows that a broad distribution of the shape index of cells affects the tissue
rigidity and, consequently, the cell diffusion coefficient~\cite{Li2019}.
Understanding the role of cell heterogeneities in the adapation time is a
question of interest for future studies.

\section{Methods}
To simulate the confluent tissue, we used a recently developed hybrid CPU/GPU
software package, cellGPU~\citep{Sussman2017}, for the self-propelled Voronoi
model. The equations of motion \eqref{motion} are integrated using the Euler
method, with a time step of $\Delta t=10^{-3}$. We impose periodic boundary
conditions, $D_r=1$, $v_0=0.3$, and $r=100$, the latter to guarantee that
fluctuations in the cell area are negligible when compared to the ones in the
perimeter. For the considered set of parameters, the rigidity transition
occurs for $p_c\approx3.725$~\cite{Bi2016}. To generate the initial
configuration, we generate $N$ positions at random and let the system relax
over $10^4$ time steps, with $p_0=p_B$ for all cells.  Then, we set
$p_{0,i}(0)=p_A$ for the cells in the left- and $p_{0,i}(0)=p_B$ for the ones
in the right-hand side of the substrate.

\section{Acknowledgments}
The authors acknowledge financial support from the Portuguese Foundation for
Science and Technology (FCT) under Contracts no. PTDC/FIS-MAC/28146/2017
(LISBOA-01-0145-FEDER-028146), UIDB/00618/2020, UIDP/00618/2020 and
SFRH/BD/131158/2017. MLM and GET acknowledge support from the Simons
Foundation (\#446222) and from NSF-DMR-1352184 (MLM).

\section{Supplementary Information}

\subsection{Reduced units}
If we rescale energy in units of $K_AA_0^2$, the energy functional $E_i$ given
by Eq.~\eqref{eq::en.functional} is,
\begin{equation}\label{eq::en_func_reduced}
e_i=(a_i-1)^2+\frac{(p_i-p_0)^2}{r} \ \ ,
\end{equation}
where $e_i=E_i/K_AA_0^2$, $a_i=A_i/A_0$ and $p_i=P_i/\sqrt{A_0}$ are the area and
perimeter of cell $i$ in adimensional units, $p_0=P_0/\sqrt{A_0}$ is the shape
index, and the ratio $r=K_AA_0/K_p$ sets the ratio between the area and the
perimeter moduli.

\subsection{Influence of $\epsilon$ on the results}
In Eq.~\eqref{DP}, the definition of alike cells depends on a threshold
$\epsilon$, which we fixed at $\epsilon=10^{-5}$. In practice, the value of
these $\epsilon$ will depend on the experimental resolution to segregate cells
by their type. Here, we study the dependence on $\epsilon$. 

\begin{figure}[t]
\includegraphics[width=\columnwidth]{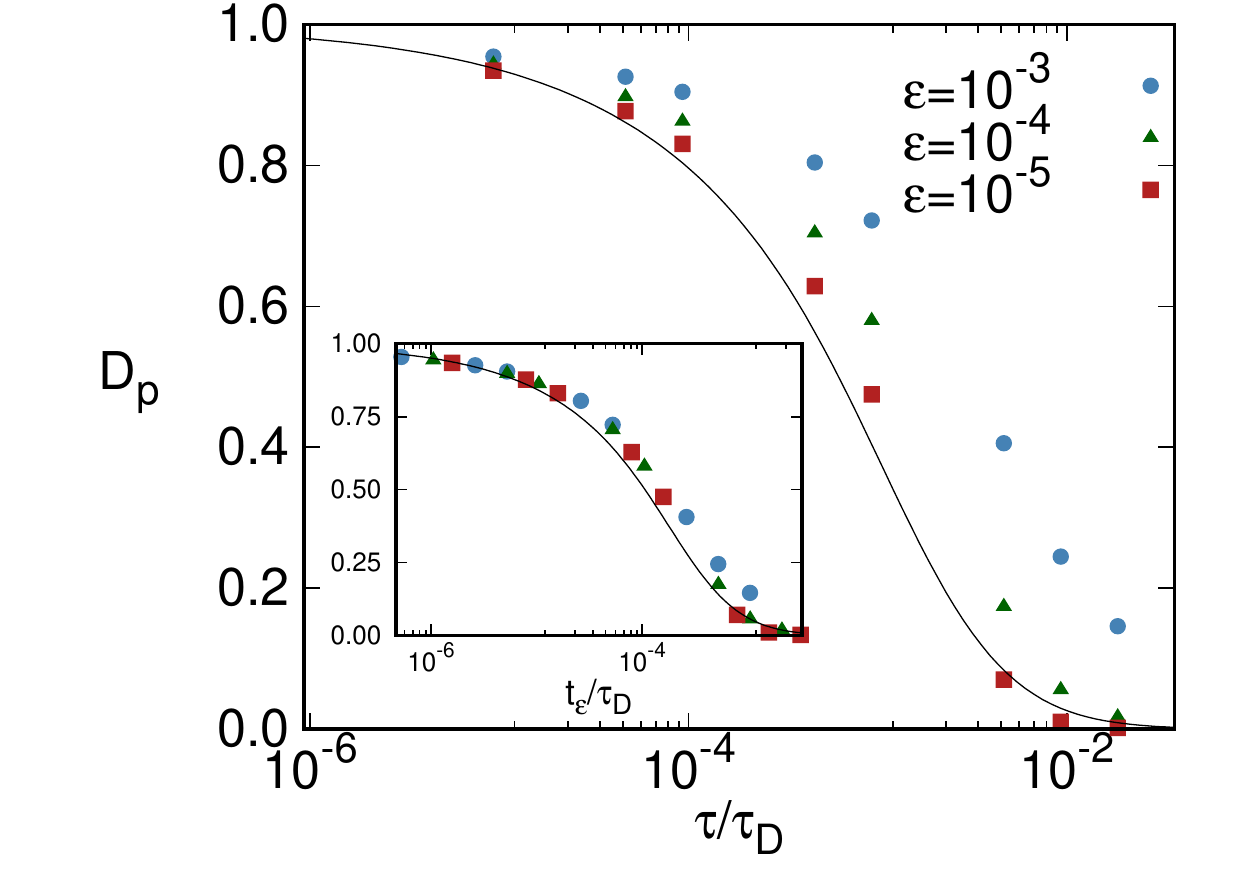}
\caption{\label{eps}Asymptotic demixing parameter $D_{p}$ as a function of
$\tau/\tau_D$, for $N=4096$. Different
curves are for different values of $\epsilon$, namely, $10^{-5}$, $10^{-4}$,
and $10^{-3}$. All the other parameters are the same as in the fluid-fluid
case of the main paper. In the inset, a data collapse is obtained when time is
rescaled as proposed in Eq.~\eqref{epsilon.eff}. Results are averages over ten
independent samples.}
\end{figure}
Figure~\ref{eps} shows the time dependence of the demixing parameter $D_p$ for
different values of $\epsilon$ for the fluid-fluid case as in the main paper.
The larger is the value of $\epsilon$, the more cells are considered to be
alike and therefore, the value of $D_p$ is larger. To compare results for
different values of $\epsilon$, we compute the time $t_\epsilon$ that takes
for a cell to adapt to a new target shape index, within a threshold
$\epsilon$. If the time evolution of the shape index is given by
Eq.~\eqref{eq::tau},
\begin{equation}\label{epsilon.eff}
t_{\epsilon}=\int_{\epsilon}^{\Delta p_{0}} -\tau \ln
\left(\frac{\epsilon}{\delta_p} \right) d\delta_p \ \ ,
\end{equation}
where $\Delta p_0=p_B-p_A$ is the change in shape index. In the inset of Fig.~\ref{eps}
we show that a data collapse is obtained for a wide range of values of
$\epsilon$ (three orders of magnitude), if we plot $D_p$ as a function of
$t_\epsilon$.

\subsection{Time dependent solution for the continuum model}
To solve Eq.~\eqref{Eq}, we define a characteristic length $L^*=L/2$
and time $T^*=L^2/4D^*$, respectively, and introduce two adimensional variables,
\begin{equation}
\hat{x}=\frac{2x}{L} \hspace{1cm} \hat{t}=\frac{4D^*t}{L^2} .
\end{equation}
Using the chain rule, we get the following identities,
\begin{equation}
u_{t}=\frac{1}{T^{*}}u_{\hat{t}} \ \ \text{and} \ \ u_{xx}=\frac{1}{(L^{*})^{2}}u_{\hat{x}\hat{x}}.
\end{equation}
By replacing them in Eq.~\eqref{Eq}, we obtain,
\begin{equation} \label{PDE}
u_{\hat{t}}(\hat{x},\hat{t})=u_{\hat{x}\hat{x}}(\hat{x},\hat{t})+\phi^2[1-u(\hat{x},\hat{t})] 
\end{equation}
where $\phi^2=TL^{2}/4D^*$. The initial and boundary conditions are then,
\begin{eqnarray}
\label{adi22}
u(0,\hat{t})=u(1,\hat{t})=0, \hspace{1cm} \hat{t}>0 , \\
\label{adi33}
u(\hat{x},0)=1, \hspace{1cm} 0<\hat{x}<1 .
\end{eqnarray}

In the results section, we present the stationary state solution
$u_E(\hat{x})$ obtained by setting $u_{\hat{t}}=0$. Here, to derive the time
dependent solution, we define,
\begin{equation}
\label{vu}
v(\hat{x}, \hat{t})=u(\hat{x},\hat{t})-u_{E}(\hat{x}).
\end{equation}
Substituting in Eq.~\eqref{PDE} gives,
\begin{equation}
v_{t}(\hat{x},\hat{t})=v_{xx}(\hat{x},\hat{t})-\phi^2v(\hat{x},\hat{t}) \ \
,
\end{equation}
and the boundary conditions are now,
\begin{eqnarray}
\label{adiv2}
v(0,\hat{t})=v(1,\hat{t})=0, \hspace{1cm} \hat{t}>0, \\
\label{adiv3}
v(\hat{x},0)=1-u_{E}(\hat{x}), \hspace{1cm} 0<\hat{x}<1.
\end{eqnarray}

This set of equations is solved by separation of variables,
$v(\hat{x},\hat{t})=X(\hat{x})T(\hat{t})$, which gives,
\begin{equation}
\frac{T'(\hat{t})}{T(\hat{t})}+\phi^2=\frac{X''(\hat{x})}{X(\hat{x})}.
\end{equation}
Imposing the initial and boundary conditions, we obtain,
\begin{widetext}
\begin{equation}
\label{eqt}
u(\hat{x},\hat{t})= \frac{4}{\pi}\sum_{n=1}^{\infty}\left\{ \left[
1-\frac{\phi^2}{(2n-1)^{2}\pi^{2}+\phi^2} \right]\frac{sin[(2n-1)\pi
\hat{x}]}{(2n-1)}e^{-[(2n-1)^{2}\pi^{2}+\phi^2]\hat{t}} \right\}+\frac{1+e^{\phi}-e^{-\phi(-1+\hat{x})}-e^{\phi\hat{x}}}{1+e^{\phi}} \ \ ,
\end{equation}
\end{widetext}
which, in the limit $\hat{t}\rightarrow\infty$ gives the stationary solution
in the main paper.

\subsection{Dependence on $\Delta p_{0}$}
\begin{figure}[t]
\includegraphics[width=\columnwidth]{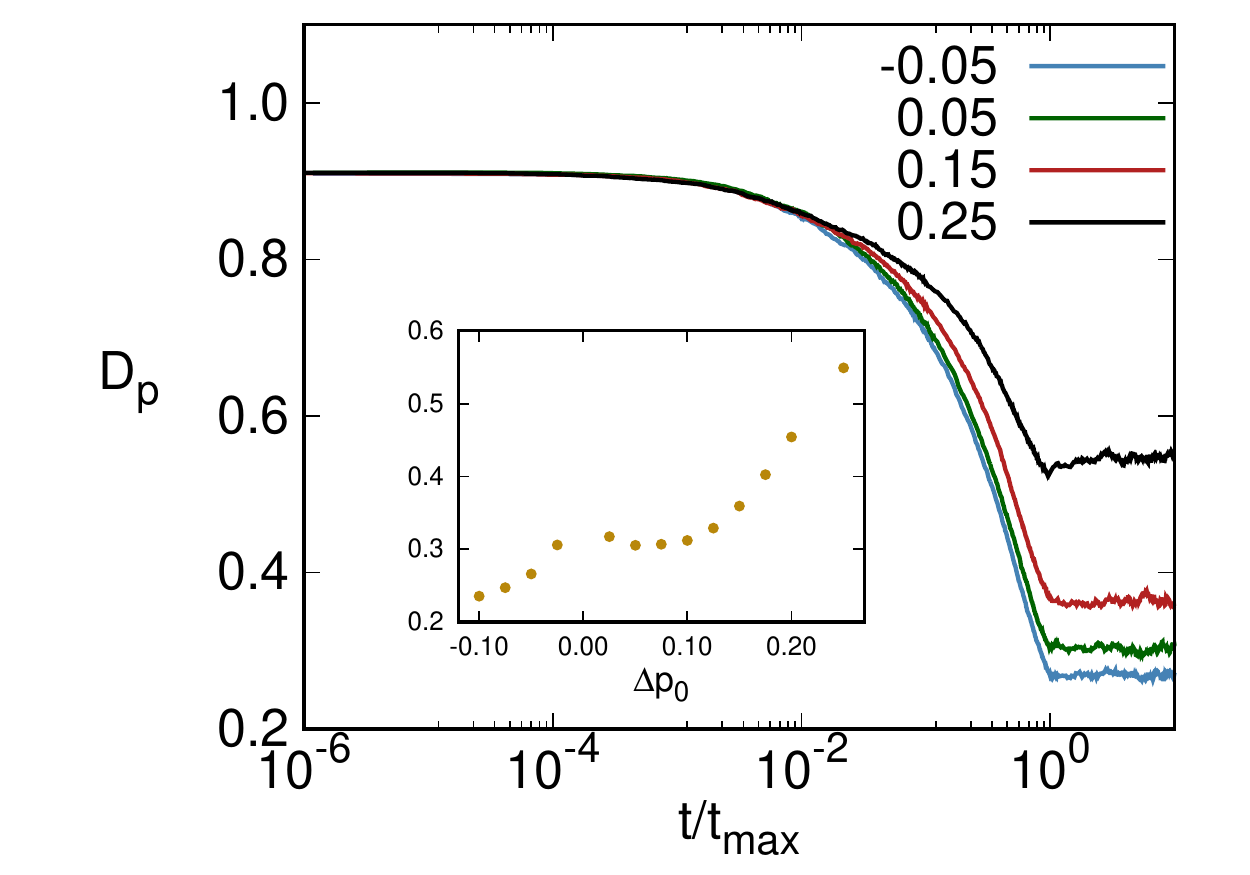}
\caption{\label{deltap0}Time dependence of the demixing parameter $D_p$, where
time is rescaled by $t_\mathrm{max}=-\ln(\epsilon/\Delta p_0)$, with $N=256$.
Different curves are for different values of $\Delta p_0=p_B-p_A$, where
$p_B=3.9$. In the inset is the asymptotic value of $D_p$ as a function of
$\Delta p_0$. Results are averages over $10^2$ samples.}
\end{figure}
As discussed in the main text, the properties of the tissue on each side of
the substrate depend, not only on the target shape index of that side but also
on the target value on the other side. Here, we explore the dependence on
$\Delta p_0=p_B-p_A$, defined as the difference in the target shape indices of
both sides. For simplicity, we fix $p_B=3.9$ and change $p_A$. 

Figure~\ref{deltap0} shows the time dependence of the demixing parameter for
different values of $\Delta p_0$. For all values, $D_p$ initially decreases
and saturates asymptotically. In the inset, we plot the asymptotic value of
$D_p$ as a function of $\Delta p_0$, which reveals a non-monotonic behavior.
From Eq.~\ref{epsilon.eff}, we see that the time it takes for a cell to fully
adapt to the new side depends on both $\tau$ and $\Delta p_0$. So, for
vanishing $\Delta p_0$, cells crossing sides swiftly adapt to the local target
shape index, leading to an increase in $D_p$. 

\bibliography{Cells}

\end{document}